\documentclass[10pt,aps,prb,twocolumn,superscriptaddress, nolongbibliography]{revtex4-2}

\usepackage[usenames,dvipsnames,svgnames]{xcolor}

\usepackage{lmodern}
\usepackage{soul}
\usepackage[T1]{fontenc}
\setcitestyle{super}

\usepackage[utf8]{inputenc}
\usepackage{graphicx}
\usepackage{xcolor}

\begin{document}

\title{Multi-Octave Frequency Comb from an Ultra-Low-Threshold Nanophotonic Parametric Oscillator}

\author{
Ryoto Sekine$^{1,*}$, Robert M. Gray$^{1,*}$, Luis Ledezma$^{1}$, Selina Zhou$^1$, Qiushi Guo$^{1,2}$, Alireza Marandi$^{1,\dagger}$\\
\textit{
$^1$Department of Electrical Engineering, California Institute of Technology, Pasadena, California 91125, USA. \\
$^2$Department of Physics, CUNY, New York, New York 10017, USA.\\
$^\ast$These authors contributed equally to this work.}\\
$^\dagger$Email: \href{mailto:marandi@caltech.edu}{marandi@caltech.edu}
}

\date{\today}

\begin{abstract}
Ultrabroadband frequency combs coherently unite distant portions of the electromagnetic spectrum. They underpin discoveries in ultrafast science and serve as the building blocks of modern photonic technologies. Despite tremendous progress in integrated sources of frequency combs, achieving multi-octave operation on chip has remained elusive mainly because of the energy demand of typical spectral broadening processes.  Here we break this barrier and demonstrate multi-octave frequency comb generation using an optical parametric oscillator (OPO) in nanophotonic lithium niobate with only femtojoules of pump energy. The energy-efficient and robust coherent spectral broadening occurs far above the oscillation threshold of the OPO and detuned from its linear synchrony with the pump. We show that the OPO can undergo a temporal self-cleaning mechanism by transitioning from an incoherent operation regime, which is typical for operation far above threshold, to an ultrabroad coherent regime, corresponding to the nonlinear phase compensating the OPO cavity detuning. Such a temporal self-cleaning mechanism and the subsequent multi-octave coherent spectrum has not been explored in previous OPO designs and features a relaxed requirement for the quality factor and relatively narrow spectral coverage of the cavity. We achieve orders of magnitude reduction in the energy requirement compared to the other techniques, confirm the coherence of the comb, and present a path towards more efficient and wider spectral broadening. Our results pave the way for ultrashort-pulse and ultrabroadband on-chip nonlinear photonic systems for numerous applications.  
\end{abstract}
\maketitle

\begin{figure*}[ht]
\centering
\includegraphics[width=0.9\linewidth]{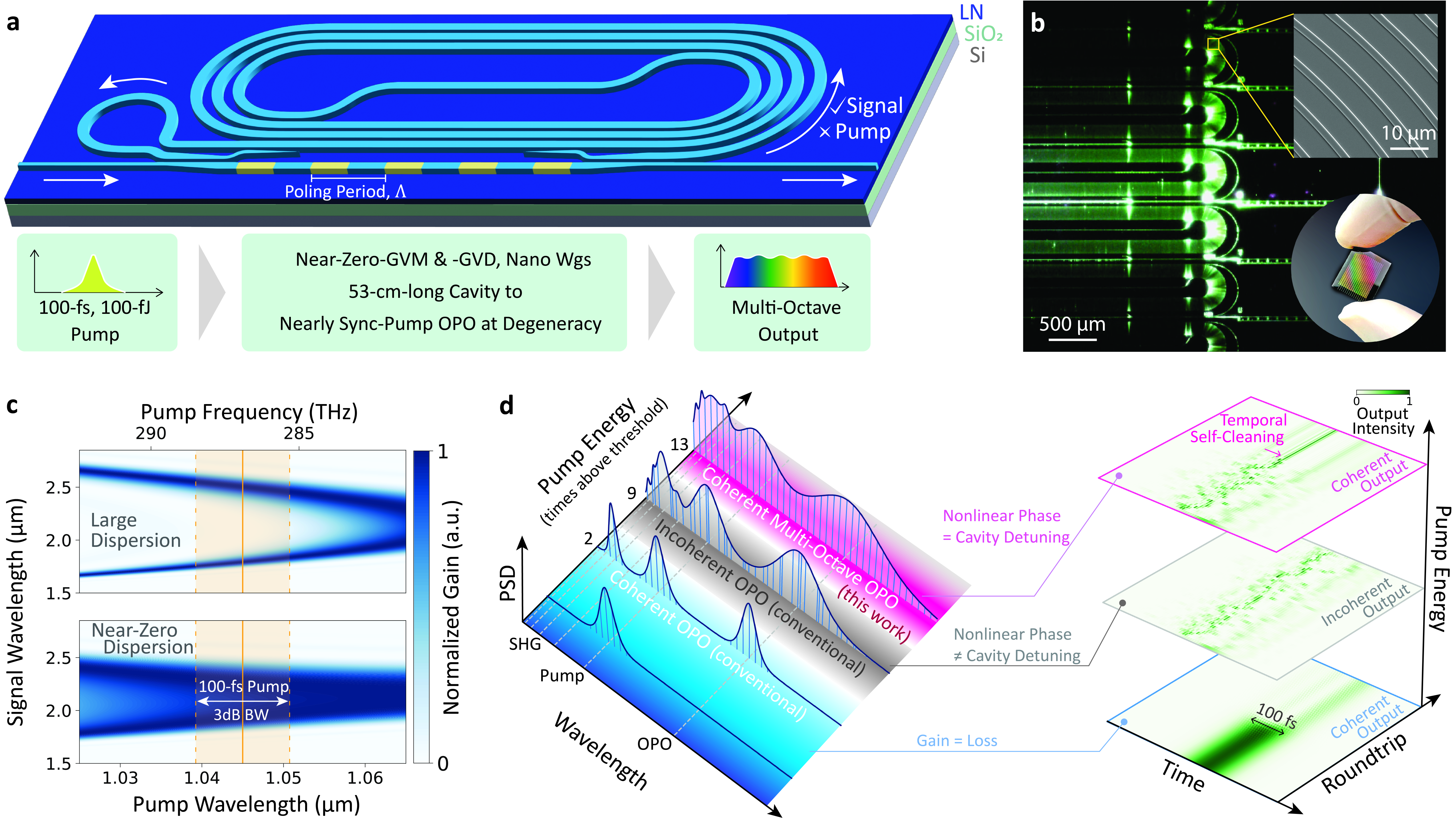}
\caption{\textbf{Principle and design of the multi-octave nanophotonic OPO}. \textbf{a,} Illustration of the sync-pumped OPO on thin-film lithium niobate with key features highlighted. \textbf{b,} Microscope image of several devices when the one in the center is pumped at 1 $\mu$m. The chip glows green due to second harmonic generation (SHG). The top inset is a scanning electron microscope image of the spiral region and the bottom is a picture of the entire chip containing 16 OPOs. \textbf{c,} Illustration showcasing how short pump pulses can take advantage of near-zero-dispersion-engineered OPAs. The simulated gain profiles are shown in the top for a waveguide with 60 fs$^2$/mm half-harmonic GVD and 26 fs/mm GVM and in the bottom for a near-zero-dispersion waveguide. The solid orange line marks the center wavelength of the pump and the orange shaded regions mark the 3-dB bandwidth (BW) of the 100-fs source. \textbf{d,} Depiction of the different regimes of operation of the OPO as a function of pump pulse energy, along with the roundtrip-to-roundtrip temporal output of the OPO in each regime. 
}\label{Fig1}
\end{figure*}

Broadband optical frequency combs are among the great achievements of modern optics \cite{Carlson2018subcycle, Scott2020freqcomb}. Recently, increasing efforts are focused on the realization of broadband frequency combs in nanophtonic platforms \cite{Kippenberg2018dks, Chang2022intofc, Gaeta2019intofc} with applications including dual-comb spectroscopy \cite{MGS2016dcs}, optical communications \cite{MarinPalomo2017comm}, optical frequency synthesis \cite{Spencer2018synthesizer, Tran2022extintfreq}, and laser ranging \cite{Riemensberger2020ranging}. However, the spectral coverage of integrated frequency comb sources remains far behind their table-top counterparts using high-pulse-energy lasers and discrete components, which have recently surpassed six-octave spectra \cite{Lesko2021sixoct, Elu2021seven}. Such multi-octave frequency combs are valuable for applications such as ultrashort pulse synthesis\cite{Wirth2011synth}, attosecond science\cite{Corkum2007ato}, and bio-chemical sensing and imaging \cite{Muraviev2018dcs, Picque2019review, Chang2003}.

Integrated sources of short-pulse frequency combs typically generate picojoules or femtojoules of pulse energies \cite{Scott2020freqcomb, Chang2022intofc, Yu2022pulses, Guo2023mll, Stern2018batterycomb} and their spectral coverage barely reaches an octave \cite{Pfeiffer2017cwdks,Li2017cwdks}. This has necessitated further spectral broadening stages for many applications, which so far have been realized strictly using table-top systems with discrete amplifiers and components \cite{Carlson2018subcycle, Obrzud2019astrocomb, Spencer2018synthesizer}. A femtojoule-level multi-octave coherent spectral broadening mechanism has so far been beyond the reach of current photonic technologies, and hence, a path towards a fully integrated multi-octave frequency comb has remained elusive.

Substantial spectral broadening is typically achieved by passing femtosecond or picosecond pulses with 0.1-10 nJ of energy through waveguides, crystals or fibers with quadratic ($\chi^{(2)}$) or Kerr ($\chi^{(3)}$) nonlinearity with various designs \cite{wu2022chirpedscg,Carlson2018subcycle,YoonOh2017uvscg,Ludwig2023uvastrocomb,Dudley2006pcf,Vasilyev2019ZnSGaSe}. Among these schemes, waveguides with quadratic nonlinearity are becoming increasingly more efficient, especially because of the recent progress on quasi-phase matching and dispersion engineering \cite{Jankowski2020scg, wu2022chirpedscg,Ludwig2023uvastrocomb} and show superior performances over their cubic counterparts. However, to reach an octave of coherent spectrum and beyond they still need 10’s of picojoules of energy\cite{Jankowski2020scg}, which is far beyond the current capability of integrated frequency comb sources. 

Resonant enhancement of spectral broadening is expected to improve the energy requirements. However, such experiments have so far remained below an octave \cite{Anderson2021pDKS, Anderson2022zdks,Obrzud2019astrocomb}. This is mainly because of the overly constrained dispersion requirements of cubic coherent spectral broadening schemes especially when combined with high-Q requirements. In fact, even linear components in nanophotonics with multi-octave spectral response are still challenging to design and realize \cite{Molesky2018inversedesign}. In contrast, quadratic nonlinearity not only leads to lower energy requirements in single-pass configurations, but it also offers a wider range of nonlinear processes for ultrawide coherent spectral broadening resulting from nonlinear interactions of distant portions of the spectrum\cite{Lesko2021sixoct, Elu2021seven}. However, a proper resonator design is necessary to enable an operation regime where a sequence of quadratic nonlinear processes can yield coherent spectral broadening towards multi-octave operation.

A promising path towards such a multi-octave nonlinear resonator is based on synchronously (sync-) pumped degenerate OPOs, which so far have been successfully used in bulk optics for efficient phase-locked down-conversion via half-harmonic generation of broadband frequency combs \cite{Muraviev2018dcs,Peter2016Ising,Marandi2012coherence,Marandi2016cascaded}. Recent studies indicate the potential of sync-pumped OPOs for extreme pulse shortening and spectral broadening while preserving the coherence properties of the pump \cite{Roy2022walkoff}. However, lack of dispersion engineering in bulk nonlinear crystals, low parametric gain bandwidths, and multi-picojoule thresholds have put limitations on their applicability for compact and ultrabroadband frequency comb applications. Recent developments of dispersion-engineered optical parametric amplifiers (OPAs) \cite{Ledezma2022OPA} and narrowband sync-pumped OPOs \cite{Roy2022SyncOPO} in lithium niobate nanophotonics promise a path towards overcoming these limitations and accessing a new regime of ultrabroadband ultra-low-energy nonlinear optics that has not been accessible before.  

\begin{figure*}[ht]
\centering
\includegraphics[width=\linewidth]{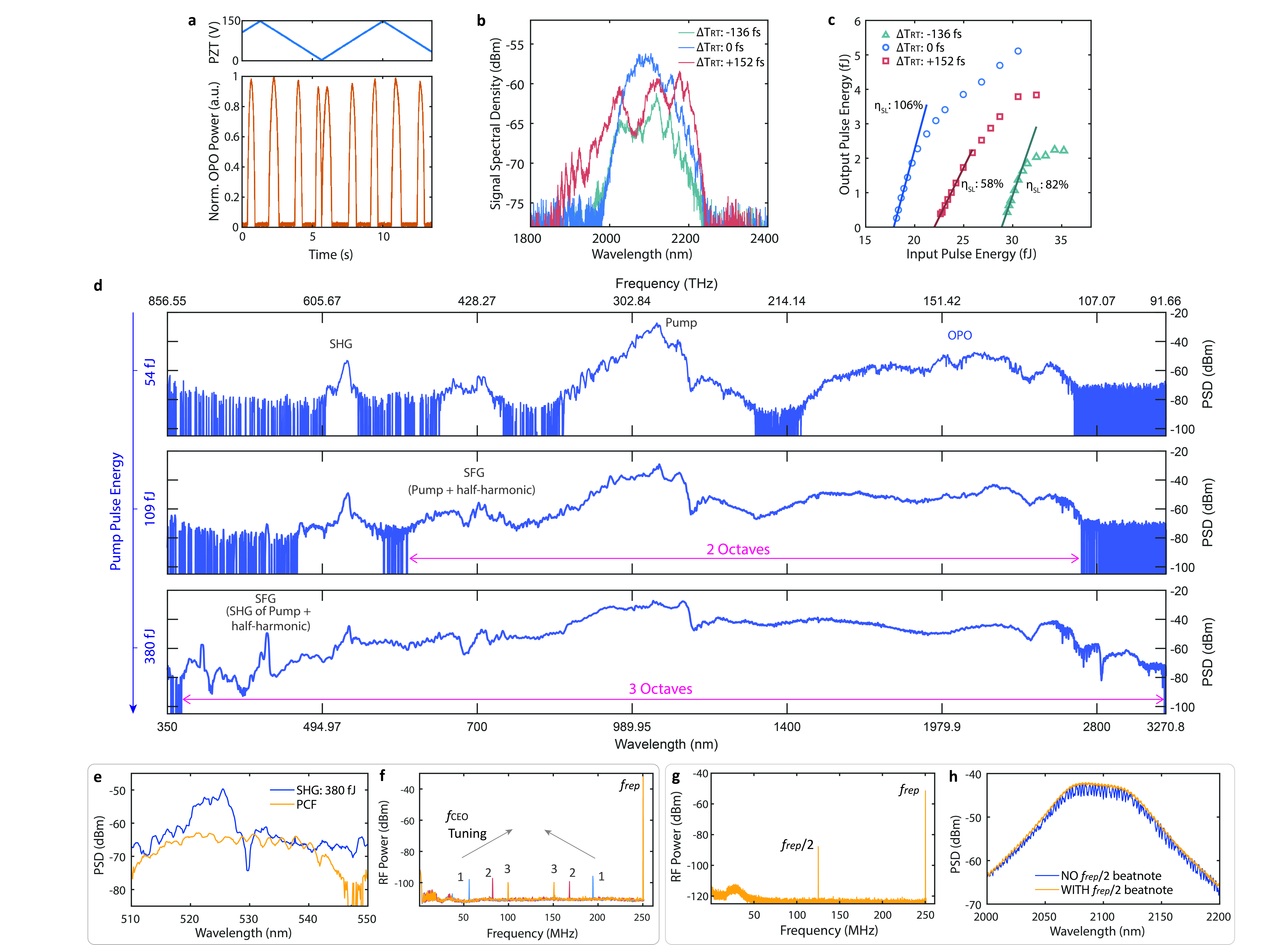}   
\caption{\textbf{OPO characterization.} 
\textbf{a,} Oscillation peaks of the OPO as the pump repetition rate is modulated by a piezoelectric transducer (PZT) in the pump laser cavity at 600 Hz. \textbf{b,} Signal spectrum at 35 fJ of pump energy for three different roundtrip detunings and \textbf{c,} the corresponding OPO signal growth as a function of pump energy for different oscillation peaks and their slope efficiencies, $\eta_\mathrm{SL}$. \textbf{d}, Output spectra from the OPO cavity at 54 fJ, 109 fJ, and 380 fJ of pump. \textbf{e}, Spectral overlap between the broadened pump using a PCF and the SHG output of the integrated device with a cavity. \textbf{f}, The resulting radio-frequency (RF) beatnote at three different pump $f_\mathrm{CEO}$ frequencies when the chip is pumped at 380 fJ. \textbf{g}, Example beatnote between a free space and on-chip OPO pumped at 109 fJ and $\sim$8 fs of cavity detuning. \textbf{h}, Interference of these two coherent OPOs filtered around 2.1 $\mu$m as their relative delay is scanned. The interference is observed only when their $f_\mathrm{CEO}$ frequencies are the same, while the beatnote is observed only when they are different by $f_\mathrm{rep}$/2.
}\label{Fig2}
\end{figure*}

In this work, in sharp contrast to previous realizations of nonlinear photonic resonators, we judiciously design and realize an on-chip sync-pumped OPO featuring a low-finesse resonator which couples only frequencies near the half-harmonic of the pump while leaving the pump and its high-harmonics non-resonant. It is near-zero dispersion engineered for the pump and its half-harmonic. The nanophotonic sync-pumped OPO operates with a record-low threshold of $\sim$18 fJ. Due to its low-energy, intense, phase-sensitive amplification, we discovered an operation regime of the OPO where the nonlinear phase compensates the cavity detuning, yielding temporal self cleaning and a multi-octave coherent spectrum.  We measured a two-octave frequency comb at $\sim$109 fJ of pump energy and experimentally confirmed its coherence. We numerically replicate the broadband nonlinear dynamics associated with such a multi-octave broadening and provide design guidelines for even broader outputs.

\begin{figure*}[ht]
\centering
\includegraphics[width=0.9\linewidth]{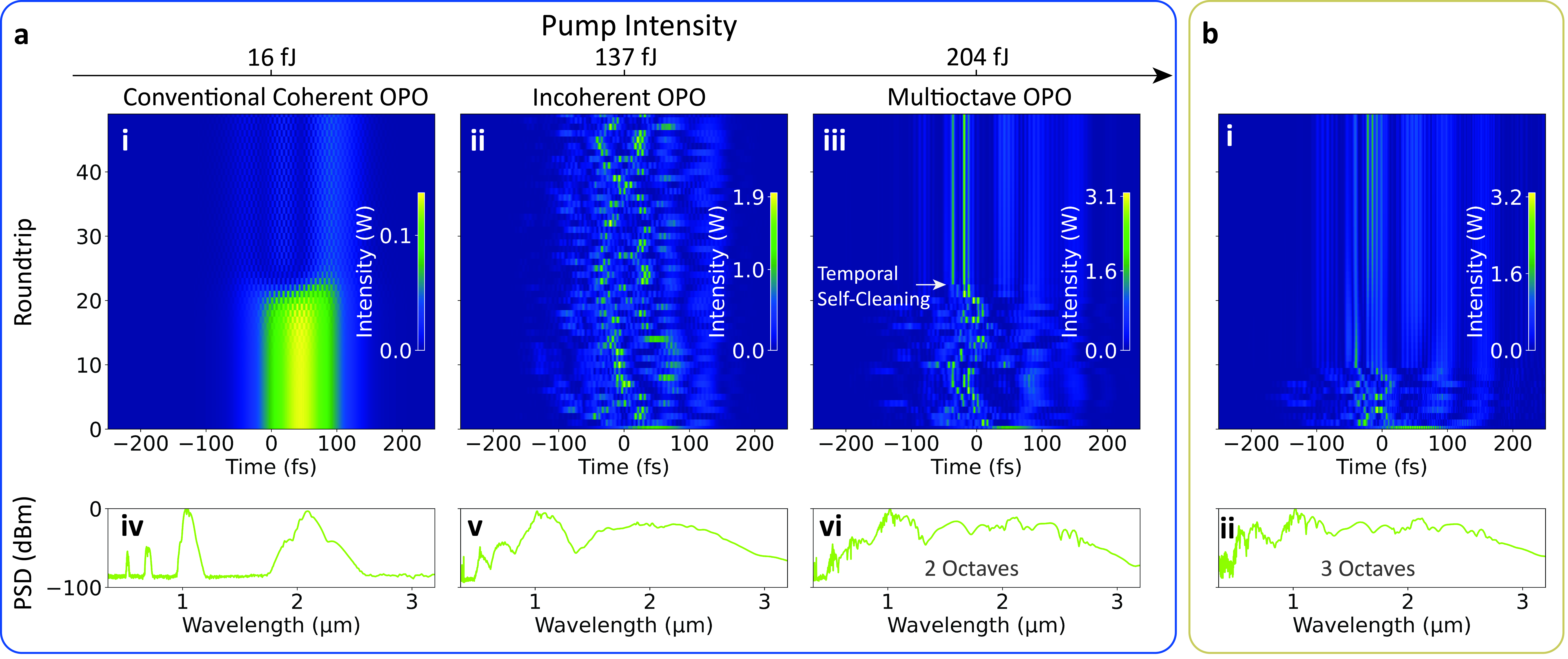}
\caption{\textbf{Simulation results showing different operation regimes of the nanophotonic OPO}. 
\textbf{a}, Transition from (i) near-threshold coherent operation to (ii) incoherent operation and (iii) back to coherent operation when the pump energy is increased. The roundtrip temporal evolution (i-iii) and output spectra (iv-vi) are shown for three different pump intensities using experimental parameters and at a cavity detuning of -10.5 fs. \textbf{b}, A three-octave coherent OPO. The same experimental parameters are used except that the last one mm of the PPLN was replaced with a chirped poling period. The pump pulse energy was at 250 fJ.
}\label{Fig3}
\end{figure*}

\textbf{Operating principle and design}

Figure \ref{Fig1}a illustrates the design of the on-chip sync-pumped OPO, with the fabricated device shown in Fig. \ref{Fig1}b. The input/output couplers are designed to allow resonance only around the half-harmonic of the pump (see supplementary section I), and the cavity is designed to be minimally dispersive for these wavelengths. To phase and frequency lock the OPO, the OPO is nearly sync-pumped at degeneracy, requiring a cavity round-trip time of 4 ns for a pump comb with a 250 MHz repetition rate. With the effective index of our nanophotonic lithium niobate waveguides (wgs), this amounts to a 53-cm-long-cavity.

To achieve the ultra-high, ultra-broad, phase-sensitive gain at fJ pump pulse energies that enables coherent broadband comb generation, the OPO includes a 10.8 mm OPA with proper dispersion engineering and quasi phase matching (QPM). Specifically, we target minimizing the group velocity dispersion (GVD) of the pump and signal, as well as the group velocity mismatch (GVM) between the pump and signal \cite{Ledezma2022OPA}. Figure \ref{Fig1}d illustrates the large gain bandwidth that can be accessed when coupling a 100-fs pump to a near-zero dispersion engineered waveguide, as opposed to one with large dispersion that is favored for broadly tunable OPOs\cite{Ledezma2022CWopo, Roy2022SyncOPO}. The designs for the poling period, cavity length, and couplers for sync-pumped operation can be found in the Supplementary, Section I.

Figure \ref{Fig1}d illustrates the different regimes of operation of this nanophotonic OPO. At low pump pulse energies, the OPO goes above threshold when the gain overcomes the loss inside the cavity. This is conventionally the regime where OPOs are operated to yield coherent outputs phase-locked to the pump\cite{Marandi2012coherence}. At higher pump pulse energies a degenerate OPO is known to transition to an unstable operation regime where the phase-locked operation deminishes\cite{Ryan2016opo,Mosca2018miopo}. Here however, we find that far above threshold, the OPO can undergo a transition to the phase-locked regime as a result of the nonlinear phase being compensated by the cavity. This emergence of coherence is akin to the spatial self-cleaning in multimode fibers\cite{Krupa2017selfcleaning}, which is emphasized in the accompanying time domain plots as a temporal self-cleaning mechanism, where after a finite number of roundtrips the output pulse intensity is seen to stabilize with ultrashort features in the multi-octave case.


\textbf{Experimental results}

In Fig. \ref{Fig2}a-c, we show the near-threshold performance of the nanophotonic OPO. Scanning the repetition rate of the pump by 600 Hz, we observe the oscillation peaks of the OPO as depicted in Fig. \ref{Fig2}a. These peaks are characteristic of doubly-resonant operation\cite{Marandi2012coherence}. We can actively lock the pump repetition rate to the center of each of these peaks, and the near-threshold signal spectra of three such peaks at distinct detunigs between the pump repetition period and cavity round-trip time, $\Delta T_{RT}$, are shown in Fig. \ref{Fig2}b. In Fig. \ref{Fig2}c we show the measured input-output pulse energy growth of these same peaks. We can extrapolate the threshold and slope efficiencies, $\eta_{SL}$, and define the peak with the lowest threshold as the zero cavity detuned state. For this peak we estimate an OPO threshold of $\sim$18 fJ.

In Fig. \ref{Fig2}d, we show three characteristic output spectra of the OPO. At 54 fJ of pump we observe conventional OPO behavior. The pump, half-harmonic and second-harmonic are all spectrally broadened, and there is noticeable sum frequency generation (SFG) between the pump and half harmonic. At 109 fJ of pump, we observe continuous spectra from 600 nm to 2710 nm, and at 380 fJ, we observe three-octave-spanning spectra from 362 nm to 3261 nm. Notably for the spectra above 2.5 $\mu$m, we observe molecular absorption features, here predominately due to ambient H$_2$O. The dip at 2.8 $\mu$m is associated with the OH absorption peak in the LN and/or the buffer layer\cite{Ledezma2022CWopo,Leidinger2015LNabs}, and kinks near 680 nm and 1135 nm are due to mode crossings (see Supplementary Section II).

\begin{figure*}[ht]
\centering
\includegraphics[width=0.9\linewidth]{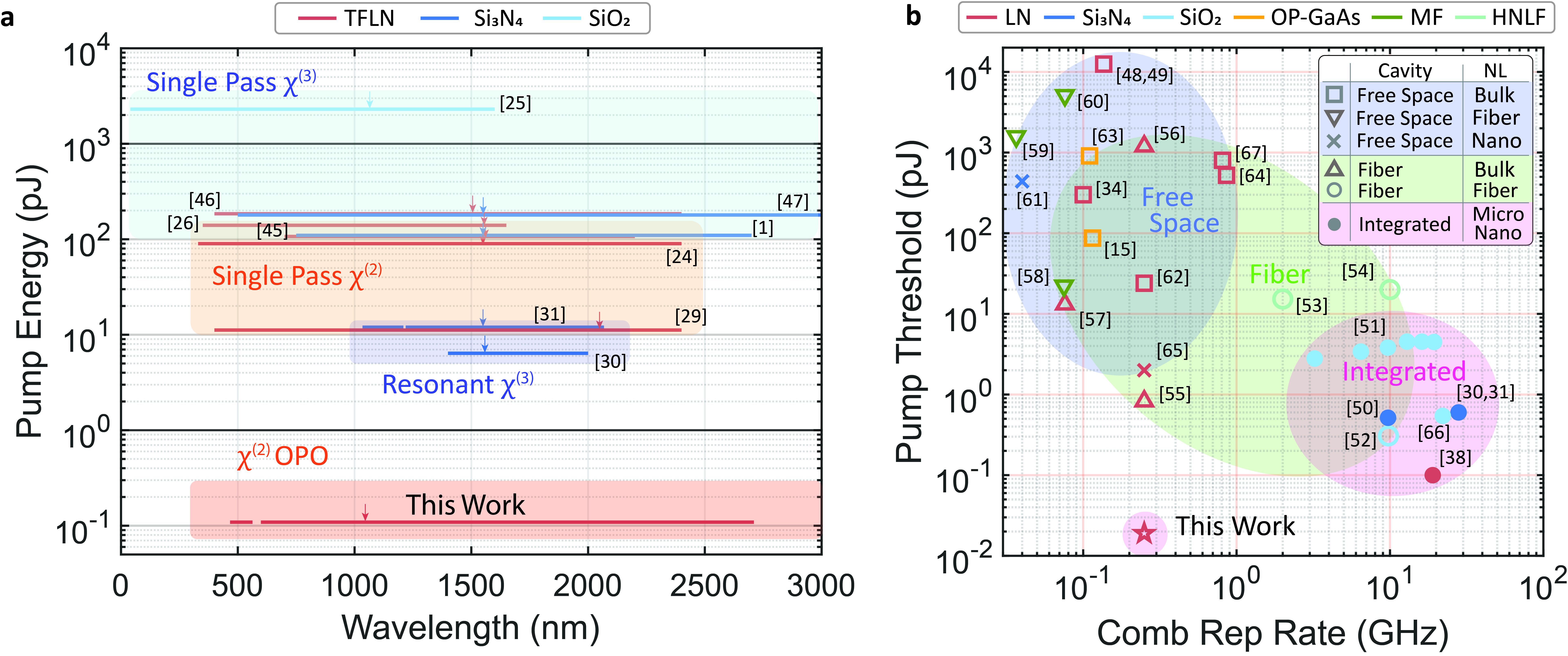}
\caption{\textbf{Performance comparison of (a), integrated spectral broadening, and (b), frequency comb sync-pumped OPOs}. \textbf{a}, Wavelength coverage and pump pulse energies of integrated frequency comb spectral broadening schemes. The arrows indicate the pump wavelength. \textbf{b}, Comb repetition rates and pump threshold energies of sync-pumped OPOs. The marker shapes denote the different cavity and nonlinear (NL) element compositions for each OPO, the categories being free space, fiber, integrated and bulk, fiber, nanophotonic respectively. In both figures, the top legend denotes the material of the nonlinear element. Abbreviations, TFLN: thin-film lithium niobate, OP: orientation patterned, MF: microstructured fiber, HNLF: highly nonlinear fiber.
}\label{Fig4}
\end{figure*}

First, we investigate the coherence of the second-harmonic portions of these spectra using a spectrally broadened output of the pump by a photonic crystal fiber. We interfere this broadened pump with the second-harmonic portion of the on-chip OPO, the spectral overlap of which is shown in Fig. \ref{Fig2}e. We show sample beatnotes of the resultant carrier-envelope offset frequency, $f_\mathrm{CEO}$, along with the pump repetition rate, $f_\mathrm{rep}$, at 250 MHz when pumping at 380 fJ in Fig. \ref{Fig2}f. By observing the shifting of these beatnotes as the pump $f_\mathrm{CEO}$ is tuned, we verify that these beatnotes correspond to $f_\mathrm{CEO}$. We observe similar $f_\mathrm{CEO}$ beatnotes at 54 and 109 fJ of pump (see Supplementary Section II). We also note that, as expected, these beatnotes were present irrespective of the exact detuning of the cavity for all three pump pulse energies.

To investigate the coherence of the longer-wavelength side of the on-chip OPO output, we interfere it with that of a free-space OPO pumped by the same laser using a filter centered around 2.1 $\mu$m. A degenerate OPO above threshold can have two possible CEO frequencies which differ by $f_\mathrm{rep}/2$ depending on the oscillation peak\cite{Marandi2012coherence}. When the on-chip OPO has a different CEO from the free-space OPO, upon spatially and temporally overlapping their outputs, beatnotes at $f_\mathrm{rep}/2$ are observed in the case that the on-chip OPO is pumped at 54 fJ with zero detuning. At 109 fJ of pump energy, while we did not observe the $f_\mathrm{rep}/2$ beatnote at zero detuning, we did observe it at a cavity detuning of $\sim$8 fs, and this is shown in Fig. \ref{Fig2}g. At this same detuning, when both OPOs operate with equivalent CEO frequencies, no $f_\mathrm{rep}/2$ beatnote is observed, and as the relative delay between their outputs is scanned, the two OPOs constructively/destructively interfere resulting in the fringes in Fig. \ref{Fig2}h (see Supplementary Section III).  From these measurements, we confirm that the down-converted combs at these two pump energies are coherent with respect to the pump comb. In particular, at 109 fJ of pump, because both the half-hamonic and second harmonic combs are coherent with respect to the pump and all frequency portions of our spectrum are generated through parametric processes of these three combs\cite{Jankowski2020scg}, we conclude that at this detuning the continuous two-octave wide spectrum as well as the second harmonic comb are coherent. However, for the case of 380 fJ pumping, the beatnote and spectral fringes of Fig. \ref{Fig2}g and h were not observed for any roundtrip detuning, and hence we consider this three-octave spectrum incoherent.

To explain the dynamics of this OPO far above threshold and how coherence can be established over such a broad spectrum, we turn to numerical simulations. To capture the multi-octave nonlinear interactions occurring in the OPO, we modeled the electric field in the nanophotonic cavity as a single envelope in frequency domain which is evolved using the split-step Fourier method for propagation in the PPLN region and a linear filter for the cavity feedback (see Supplementary Section III for details). In Fig. \ref{Fig3}a, we show how this captures distinct regimes of operation when using parameters matching that of the experiment. At 16 fJ the OPO goes above threshold and stabilizes after $\sim$20 roundtrips. At this point, all the frequency translated components (OPO, SHG, SFG of the pump and OPO) are coherent with respect to the pump and they remain unchanged from roundtrip to roundtrip. As the pump pulse energy is increased, fewer roundtrips are required for the OPO to form, and at 137 fJ of pump ($\sim$9$\times$ above threshold) we see that the OPO output is incoherent. 

At roughly 204 fJ of pump ($\sim$13$\times$ above threshold), however, the the half-harmonic is seen to acquire a $\pi$ phase shift through the nonlinear interaction with the pump in each single-pass through the PPLN region. This can be compensated by detuning the cavity by an odd number of OPO peaks, or by adding a constant phase offset of $\pi$ between the pump and cavity, corresponding to the carrier-envelope offset phase, $\phi_\mathrm{CEO}$, of the pump (see Supplementary Section III). The former case is shown in Fig. \ref{Fig3}a(iii) and shows a two octave coherent continuous comb that stabilizes after roughly twenty roundtrips with temporal features as short as 4 fs (see Supplementary Section III). The output spectrum is also very similar to the detuned 109 fJ experimental result of Fig. \ref{Fig2}d. 

In simulation, we further investigate how to extend the coherent operation of the OPO to even broader spectra. By replacing the last one mm of the PPLN region with a chirped poling period for efficient second harmonic and sum-frequency generation, we achieve a coherent three-octave continuous frequency comb with $\sim$250 fJ of pump energy as shown in Fig. \ref{Fig3}b.

\textbf{Conclusion and discussion} 

In Fig. \ref{Fig4} we compare our results with other integrated spectral broadening schemes and sync-pumped OPOs. The figure highlights how our nanophotonic OPO design and its operation regime enable orders-of-magnitude improvement in the energy efficiency of coherent spectral broadening. Our work represents the lowest threshold sync-pumped OPO which is enabled by its near-zero-dispersion design. This ultralow-threshold operation enabled accessing a previously unexplored operation regime of the OPO far above threshold, where ultrabroad coherent spectral broadening is established as a consequence of the balance between cavity detuning and nonlinear phase shift.

In summary, we have experimentally demonstrated a nearly sync-pumped nanophotonic OPO operating in the near zero-GVM, zero-GVD, fs-pumped, high-gain low-finesse regime resulting in an ultra-broadband coherent output with only $\sim$109 fJ of energy. The two-octave frequency comb enables unprecedented opportunities for on-chip applications including wavelength division multiplexing\cite{MarinPalomo2017comm}, dual-comb spectroscopy\cite{Picque2019freqcomb}, and frequency synthesis\cite{Gaeta2019intofc}. We show the OPO transitions from an incoherent to coherent operation regime and demonstrate a path towards much broader frequency comb sources in the femtojoule regime.

\section*{Methods}
\noindent \textbf{Device fabrication.} Our device was fabricated on 700-nm-thick X-cut MgO-doped thin-film lithium nioabte on a SiO$_2$/Si substrate (NANOLN). Following the procedure in \cite{Ledezma2022OPA}, we pattern Cr/Au poling electrodes with 16 fixed poling periods ranging from 4.955-5.18 $\mu$m using lift-off and and apply a voltage to periodically flip the ferroelectric domains. Upon poling we remove the electrodes and subsequently etch the waveguides using Ar-milling and Hyrdogen Silsesquioxane (HSQ) as the etch mask. Finally, the waveguide facets are mechanically polished to allow for butt coupling. Each OPO has a footprint of 0.5 mm $\times$ 13 mm. \\ 

\noindent \textbf{Optical measurements.} The measurements were performed using a Menlo Orange HP10 Yb mode-locked laser (MLL) centered at 1045 nm. It outputs 100-fs-long pulses at 250 MHz with a $\pm1$ MHz tuning range. Light was coupled to and from the chip using Newport 50102-02 reflective objectives, chosen for their minimal chromatic aberration. All of the results in this paper were performed on a device with 5.075 $\mu$m poling period at 26\textdegree C, regulated by a thermoelectric cooler (TEC). The lowest OPO threshold was obtained from a pump repetition rate of 250.1775 MHz, which we define as the zero detuned state. This device has a total throughput loss of 43.4 dB, and following the methodology in \cite{Ledezma2022OPA}, we measured the input and output coupling losses to be 35.7 dB and 7.7 dB respectively. For the results in Fig. \ref{Fig3}a, the spectra were collected by two different optical spectrum analyzers (OSA), specifically a Yokogawa AQ6374 (350-1750nm) and AQ6376 (1500-3400 nm). For the three octave spectra the wavelengths below 700 nm were taken with a high OH silica fiber (Thorlabs M133) and the spectra above 1750 with an InF$_3$ fiber (Thorlabs MF12). All the other measurements were taken using a low OH silica fiber (Thorlabs M72). The RF spectra was collected by an electronic spectrum analyzer (Rhode \& Schwarz FSW), combined with a high-speed silicon avalanche photodiode (Menlo Systems APD210) in Fig. \ref{Fig2}f and an InGaAs high speed photodiode (DSC2-40S) in Fig. \ref{Fig2}g.\\ 

\noindent \textbf{Numerical simulations.} We used commercial software (Lumerical Inc.) to solve for the waveguide modes shown in Sections I and II of the Supplementary that allowed us to dispersion engineer and quasi-phase-match our device. For the nonlinear optical simulation, we solved an analytical nonlinear envelope equation as described in Section III of the Supplementary. The simulations were performed with no constant phase offset between the pump and cavity unless specifically mentioned otherwise. This parameter effectively acts as a carrier-envelope offset phase of the pump, $\phi_\mathrm{CEO}$. As the simulations were performed with a time window of 1.7 ps, it should be mentioned that a large portion of the short wavelength side of the spectrum walked out of the time window of our simulation. For example, the simulated GVM between our simulation reference frame at the half-harmonic signal wavelength of 2090 nm and the second harmonic of the pump at 522 nm is 721 fs/mm. As a result, the up-converted portions of the spectrum in simulation tend to be smaller than what was measured experimentally. In these simulations we have only incorporated the effects of $\chi^{(2)}$ nonlinearity and have not considered the effects of $\chi^{(3)}$. Especially given the low pulse energies and low-finesse nature of our cavity, we believe this to be a good approximation, yet it could be one additional reason for small discrepancies between experiment and simulation.

\section*{Data Availability}
The data that support the plots within this paper and other findings of this study are available from the corresponding author upon reasonable request.

\section*{Code Availability}
The computer code used to perform the nonlinear simulations in this paper is available from the corresponding author upon reasonable request.

\section*{Acknowledgements}
The device nanofabrication was performed at the Kavli Nanoscience Institute (KNI) at Caltech. The authors thank Dr. Mahmood Bagheri for loaning equipment. The authors gratefully acknowledge support from ARO grant no. W911NF-23-1-0048, NSF grant no. 1846273 and 1918549, AFOSR award FA9550-20-1-0040, the center for sensing to intelligence at Caltech, and NASA/JPL. The authors wish to thank NTT Research for their financial support.

\section*{Authors Contributions}
R.S. and A.M. conceived the project; R.S. fabricated the devices with assistance from L.L., S.Z, and Q.G. R.G. and R.S. performed the measurements and R.G. carried out the simulations with initial input from L.L. R.S. and A.M. wrote the manuscript with inputs from all authors. A.M. supervised the project.

\section*{Competing Interests}
L.L., and A.M. are inventors on granted U.S. patent 11,226,538 that covers thin-film optical parametric oscillators. R.S., R.G., L.L., A.R., and A.M. are inventors on a U.S. provisional patent application filed by the California Institute of Technology (application number 63/466,188). R.G., L.L., and A.M. are inventors on a U.S. provisional patent application filed by the California Institute of Technology (application number 63/434,015) on 20 December 2022. L.L. and A.M. are involved in developing photonic integrated nonlinear circuits at PINC Technologies Inc. L.L. and A.M. have an equity interest in PINC Technologies Inc. The other authors declare that they have no competing interests.

\nocite{Okawachi2020f2f,Yu2019scg,Carlson2017scg}

\nocite{Spaun2016molopo,Bjork2016freqcomb,Weng2021gainswitch,Xu2020divN,Obrzud2017fmc,Inagaki2016ising,Devgan2005fibsol,Roy2023spt,Ingold2015ff,Langrock2007ffsync,Sharping2002ff,Deng2005ff,Sharping2007ff,Gao2022wgfs,Marandi2014Ising,Muraviev2018dcs,Heckl2016opgas,Reid1998fopo,Sekine2022hybrid,Li2022pdks,Burr1997opo}

\newpage
\bibliography{references}

\end{document}